# Searching for Carrington-like events and their signatures and triggers

Elena Saiz[1], Antonio Guerrero[1], Consuelo Cid[1], Judith Palacios[1] and Yolanda Cerrato[1]

[1]Space Research Group - Space Weather, Departamento de Física y Matemáticas, Universidad de Alcalá, 28871 Alcalá de Henares, Spain.

Corresponding author: Elena Saiz, Space Research Group - Space Weather, Departamento de Física y Matemáticas, Universidad de Alcalá, E28871 Alcalá de Henares (Madrid), SPAIN (elena.saiz@uah.es).

**Abstract.** The Carrington storm in 1859 is considered to be the major geomagnetic disturbance related to solar activity. In a recent paper, *Cid et al.* [2015] discovered a geomagnetic disturbance case with a profile extraordinarily similar to the disturbance of the Carrington event at Colaba, but at a mid-latitude observatory, leading to a re-interpretation of the 1859 event. Based on those results, this paper performs a deep search for other 'Carrington-like' events and analyses interplanetary observations leading to the ground disturbances which emerged from the systematic analysis. The results of this study based on two Carrington-like events (1) reinforce the awareness about the possibility of missing hazardous space weather events as the large *H*-spike recorded at Colaba by using global geomagnetic indices, (2) argue against the role of the ring current as the major current involved in Carrington-like events, leaving field-aligned currents (FACs) as the main current involved, and (3) propose abrupt southward reversals of IMF along with high solar wind pressure as the interplanetary trigger of a Carrington-like event.

1. Introduction

The prevailing paradigm assumes that when a solar ejection turns the IMF southward, remaining southward for a prolonged period of time, results in a geomagnetic storm, one feature of which is a decrease in the surface magnetic field of the Earth [e.g. *Burton et al.,* 1975; *Tsurutani and Gonzalez*, 1995]. On 2 September 1859, the Colaba observatory (10º MLat, 147º MLong) measured the most extreme geomagnetic disturbance ever recorded at low and mid latitudes, the so-called Carrington event (e.g. *Cliver and Dietrich*, 2013]). However, due to the absence of interplanetary data at that time the role of southward IMF $B_z$ in triggering the storm is unknown. As a result, a large amount of guesswork has been done concerning the Carrington storm (see as an example *Tsurutani et al.* [2003], [2005], *Siscoe et al.* [2006], and *Manchester et al.* [2006] concerning the identification of the interplanetary trigger).

*Cid et al.* [2015] found that the profile of the horizontal component ($H$) of the terrestrial magnetic field measured in a two-day period extending from 16:00 MLT, 28 October 2003 to 16:00 MLT, 30 October 2003 at Tihany magnetic observatory (in Hungary) was very similar to that recorded at Colaba during the same MLTs on 1-3 September 1859. Although the intensity of the disturbance was just half of that recorded at Colaba, many other similarities between those events guided *Cid et al.* [2015] to label the event recorded at Tihany as a 'Carrington-like' event.

A detailed analysis of magnetic records at different locations during the event on 29 October 2003, when a large dataset was available from modern terrestrial surface observatories, provided very interesting results that, extrapolated to the Carrington event, led *Cid et al.* [2015] to conclude that the *Dst* or *SYM-H* indices might have missed the large $H$-spike recorded at Colaba and to suggest that field-aligned currents (FACs) played a major role in the $H$-spike in Carrington-like events. These conclusions shook some previous results on Carrington storm, which considered that the negative $H$-spike was a drop of *Dst* and therefore related to a ring current enhancement as a consequence of a long duration and intense southern $B_z$, due to a magnetic cloud [*Tsurutani et al.*, 2003; *Li et al.*, 2006]. Besides the $H$-spike at Colaba for the 1859 event had also been suggested as not attributed to the ring current by some authors (e.g. *Green and Boardsen* [2006]), it should be noted that only extremely large IMF $B_z$ component has been associated with the terrestrial disturbance (e.g. *Ngwira et al.* [2014] and references therein).

The goal of this paper is to discover other 'Carrington-like' events in local magnetograms and, after analysing their interplanetary triggers, to extrapolate the results to the Carrington storm at Colaba. Section 2 shows the result of a deep search for other 'Carrington-like' events in the period when interplanetary data are available. Section 3 describes geospace disturbances during the Carrington-like event on 21 January 2005. In Section 4 we analyse solar and interplanetary observations of the two Carrington-like events observed at mid latitude which emerged from the systematic analysis. Finally, Section 5 is dedicated to discussion and Section 6 to conclusions.

## 2. Searching for 'Carrington-like' events in local magnetic records

We have implemented a pattern recognition algorithm [*Theodoridis et al.*, 2010] as an automated search engine to identify Carrington-like storms which is labeled CLSE (**C**arrington **L**ike **S**earch **E**ngine). The procedure let us quantify similarities between ground magnetic signatures of any event at any observatory with a previously established pattern or model. At this stage we have constrained the model pattern for similarity measure to the Carrington event (Model-C, 'Carrington').

The events subject to be compared with the model appear as a result of a systematic analysis of a subset of data obtained from the INTERMAGNET database. The subset includes data on the period from 1991 to 2009 from nine observatories (each of them representing a latitude and longitude in a three by three worldwide sector grid): ABG, KAK, and SJG for low latitude; THY, BSL, and SPT for mid latitude; and ABK, FCC, and NAQ for high latitude (see Table 1 for geographical and magnetic coordinates of the observatories).

As described above, Model-C is the Carrington geomagnetic event at Colaba observatory recorded in 1859 (hereafter C59) and it is used as the model pattern input for the comparison in the CLSE. The first task to be done is to prepare Model-C data with the same temporal resolution as the events to be compared, i.e., to get C59 1-min resolution data to be compared to other 1-min resolution data from the INTERMAGNET database. Therefore in order to get C59 data ready for the algorithm we first digitized data from Figure 3 of *Tsurutani et al*. [2003] (which displays data for a two-day interval (1 September 16 h to 3 September 16 h Bombay local time), and then an interpolation was done to obtain 1-min resolution data of the two-days data sample.

The second step in the procedure was to obtain the event list to be compared to Model-C. For this purpose, the CLSE checks every 48-h time window (same interval as available for Model-C) from the subset database searching for a minimum at that time window at the same time as the Model-C minimum. This means that in that temporal window from 0 to 48 hours, the peak (minimum) value is detected at 18 hours and 43 minutes from the beginning. If this footprint is found, data of that 48h-window are saved as an event, otherwise another 48h-window (shifted 1 minute from the previous one) is checked until the complete subset database has been checked. It can be noticed that in this procedure, the time before and the time after the peak value in the 48h-window are the conditions that define the events detected by the algorithm. No other condition is imposed to select the events. For example, the intensity of the event, the UT or the MLT of occurrence of the peak value are not taken into account.

Due to the large difference between the range of values of every geomagnetic event and Model-C, normalization is required before comparison. The scaling consists in a subtraction of the maximum value in the data set and then dividing by the total range of the window as shown in Equation 1.

$$\hat{H} = \frac{H - \max(H)}{\max(H) - \min(H)} \qquad \text{Eq (1)}$$

where *H* is the value of the horizontal component of the geomagnetic field and $\hat{H}$ is the normalised value obtained. This normalization is applied to all geomagnetic events and to Model-C, so that absolute values will be lost but the profile of the events will remain.

A score value is obtained by normalizing the measure of the distance between the event and the model, dividing by 2880 (the number of minutes in two days) and subtracting from unity as shown in Equation 2.

$$Score_i(\hat{H}_i, \hat{H}_e) = 1 - \left( \frac{\sum \|\hat{H}_i - \hat{H}_e\|}{2880} \right) \qquad \text{Eq (2)}$$

where $\hat{H}_i$ and $\hat{H}_e$ are the normalized values of the horizontal component of the geomagnetic field for the model and the event respectively. Therefore, values closer to zero indicate that two data samples differ considerably while values close to one designate greater similarity between the profiles.

The highest score values obtained are highly dependent on latitude. Figure 1 shows histograms of score frequency at three observatories, FCC, THY and ABG, as representatives of high, mid and low latitudes, respectively. The total number of events emerged at each observatory is 3166 events for FCC, 3286 for THY and 2149 for ABG. The histograms show the frequency distribution with a peak value which is clearly skewed when decreasing in latitude, indicating that the probability of finding a profile like the Carrington storm (Model-C) is larger at higher latitudes.

In each latitude range, we have analysed some of the highest score events (i.e. the disturbances most similar to the Carrington storm). We find that the onset of the high-latitude events took place close to local midnight, therefore indicating substorm activity, which we have double checked by using the Substorm Database at

http://supermag.jhuapl.edu/substorms/. For the high-latitude stations, we obtain the highest score (0.94) for the event on 29 October 2003 as observed at FCC. A careful view of the stepwise drop for this event (Figure 2a) reveals a two-pulse structure.

At low-latitude observatories, the highest score corresponds to the event on 18 June 2003 at KAK, shown in Figure 2b, with a score of 0.87. When checking mid-latitude observatories, we discover two events that score the highest, the event on 29 October 2003 (Figure 2c) from THY (0.92) and the event on 21 January 2005 (Figure 2d) from BSL (0.89). These two events closely resemble Carrington-like profiles, both in duration of the negative depression around the peak and in the local time of the peak (close to 10 MLT) although their relative amplitudes are quite different when comparing with that of C59 (0.44 for the October 2003 event and 0.11 for that of January 2005).

When comparing all panels in Figure 2 it can be noticed that a score as large as 0.87 can be obtained for an event (18 June 2003 at KAK) where the *H*-spike, both its decrease and recovering, is not as sharp as that of the Carrington event at Colaba. Even more, when the fast recovery of Colaba has been also pointed out as a relevant feature to explain FACs as the cause [*Cid et al.,* 2015]. The uniqueness of the Colaba record relative to other typical storms shall be evidenced before following on, since a large score value seems to be not enough to label a storm as "Carrington-like". Therefore a threshold should be defined and any choice would be questionable. What is the minimum score value that indicates similarity?

This problem can be solved with cluster analysis (or clustering). It is a main task of exploratory data mining used in many fields, which consists in grouping a set of objects in such a way that objects in the same group (called a cluster) are more similar to each other than to those in other groups (clusters). The task can be achieved by various algorithms, including groups with small distances among the cluster members, and the appropriate clustering algorithm and parameter settings (including the distance function to use and the threshold) depend on the individual data set and applications.

In this paper, the algorithm to group storm events in two different clusters will use the distance function given by Equation 1. The two groups correspond to Model-C (C59 described above) and Model-B ('Basic'), which is conceived as a typical geomagnetic storm, where the three phases can be easily identified, as sudden commencement (SC), main and recovery phase [*Gonzalez et al.*, 1994]. Figure 3 shows a sketch of the procedure.

We chose to represent Model-B as the disturbance registered at Tihany observatory (THY) on 20 November 2003 due to its smooth and easily identifiable phases. Moreover, this is also one of the most intense storms (-422 nT as seen by *Dst* index) whose interplanetary trigger is well identified: a magnetic cloud, preceded by a sheath, with a large and long-duration $B_z$ [*Zhang et al.*, 2007]. Data for Model-B are 1-min resolution data from the INTERMAGNET database (no interpolation or smoothing is performed) from 20 November 2003 01:12 UT to 22 November 2003 01:11 UT. Model-B is obtained after scaling the sample following the same feature scaling procedure described above for Model-C. Figure 4 shows the 48 h-window which represents the normalised profile of magnetic disturbance labeled Model-B (green line). Model-C (red dot line) has been over plotted in Figure 4 for comparison purposes and to understand better the difference between both profiles (and as a result between being member of group B or C).

The distance function of Equation 1 between Model-B and Model-C, equal to 0.85, is a useful parameter which allows us to establish a threshold to classify an event in group B or C. However, scoring more than 0.85 is not enough to belong to a certain group. We will come back to this issue later. In any case, it is important to notice that this parameter will change if a different storm profile is considered as Model-B.

Once we have established Models-B and -C, the CLSE is applied to the same subset database (records from 1991-2009 from nine INTERMAGNET observatories, see above). The geomagnetic events are obtained, processed and evaluated to obtain B- and C-scores. Therefore, similarity measures are computed twice for every event detected by CLSE, one for Model-B and another for Model-C.

Classification of the events is done by comparing both scores. A scatter plot of score C versus score B for all geomagnetic events enables us to classify the events as group C (or group B) whether it is above (or below) the line of equal similarity. The farther away from that line the greater its membership to one group and not to the other. Furthermore, due to similarity between both model events, there is a limit on how much one score can differ from the other. This limit is the score that we obtain when comparing both models (equal to 0.85, see above). Therefore, we use this value as a lower threshold for the minimum score required for an event to belong to group B or C, that is, we can properly assign an event to group C when C-score $>$ 0.85 and B-score $<$ 0.85. When applying this criterion, the group C has 1443 elements at high-, 16 at mid- and 0 at low latitude.

In the same way, an event is properly assigned to group B when B-score $>$ 0.85 and C-score $<$ 0.85. It is important to note that an event with C-score $>$ 0.85 and B-score $>$ 0.85 cannot be assigned to any group, as the event is too similar to both Models. In this way, it is worth noticing that a large score value does not always mean belonging to a certain group: the larger the difference of scores (and not the larger score) the more definite the belonging to a certain group.

Regarding the observatory data, some considerations have to be noted: there are some missing data periods of one year in the INTERMAGNET database for some observatories analysed; events with data gaps were not considered, as they may raise the score giving false information, and events with spikes were also discarded for similar reasons.

The results from this analysis appear in Figure 5 separated by latitude: (a) includes results from ABG, KAK, and SJG (low latitude); (b) THY, BSL, and SPT (mid-latitude); and (c) ABK, FCC, and NAQ (high latitude). The shadowed rectangles mark areas where Carrington-like events can be located according to the established criterion. While there are no events into the shadowed area at low latitude (Figure 5a), two events, highlighted by a red circle and labeled as C03 and C05 in Figure 5b, appear clearly separated from the diagonal with a high C-score (> 0.85) at mid-latitude observatories, as those closely resembling Carrington-like profiles. These events correspond to the event on 29 October 2003 at Tihany (THY) magnetic observatory in Hungary (C-score = 0.92, B-score = 0.82) and the event on 21 January 2005 at Stennis (BSL) magnetic observatory in southern US (C-score = 0.89, B-score = 0.84). Hereafter these events at these locations will be named as C03 and C05, respectively.

The event on 29 October 2003 at FCC has also been highlighted as C03 in Figure 5c, to note that it corresponds to the same date and time. A light blue circle in Figure 5c indicates the position of the event on 21 January 2005. At high latitudes the event appears at FFC observatory but with a very poor and comparable C- (0.63) and B- (0.68) scores.

We have performed an equivalent analysis using other storms as Model-B. No other event in the sample analysed, but C03 and C05, can be properly assigned to group C.

This fact allows us to consider common features of these events as signatures of a "Carrington-like" event and the events as representative of this group.

## 3. Geospace disturbances during the event on 21 January 2005

Geospace disturbances during C03 event were analysed by *Cid et al.* [2015]. In this paper, we perform a similar analysis for C05 in order to identify whether there are more similarities between both 'Carrington-like' events than a local geomagnetic record.

Figure 6 shows the geomagnetic disturbance during the first hours of the storm event on 21 January 2005. Measurements of the horizontal magnetic field component recorded at mid-latitude (~40° MLat) observatories spread in longitude appear in Figure 6a. Panels from top to bottom are: SUA, IRT, FRN, BSL, and SPT (increasing in MLong). The number in the top right corner of every panel corresponds to the baseline value that has been subtracted to the *H* record in order to easily compare the disturbance at every observatory. To better recognize the relevance of the disturbance, we have also added in every panel the value of the difference between the maximum and the minimum value in the last 24 hours preceding the disturbance of January 21 ($\Delta H_{max\ 24h}$). This 24-hour period can be considered as quiet time since $\Delta H_{max\ 24h}$ does not exceed 55 nT at any of the five observatories.

Figure 6b shows, for the same interval of Figure 6a, the *SYM-H* and *Dst* indices (top panel), and the horizontal magnetic field component recorded at the four low-latitude observatories involved in the computation of the *Dst* index: HER, KAK, HON, and SJG. The same format as Figure 2 of *Cid et al.* [2015] has been kept to help the reader to easily compare the two events.

Regarding the *SYM-H* index, the first magnetic signature of the storm, an enhancement of approximately 50 nT corresponding to a SC (solid line S1) is observed on 21 January at 17:11 UT decreasing gradually and developing a complex profile until 18:45 UT when another sharp increase of ~50 nT is observed (solid line S2). Then, *SYM-H* index decreases gradually to a minimum value of ~91 nT at 21:15 UT, keeping this level for approximately 15 hours before starting an extraordinary slow recovery phase (not shown in Figure 6). Vertical lines showing S1 and S2 timings have been also added in Figure 6a keeping the same label. Vertical lines with labels T1 and T2 included in Figure 6 are explained below.

In Figure 6a we compare the measurements from a set of magnetic observatories at similar geomagnetic latitude (~40º) and spread in longitude, keeping our attention in the first two hours after the SC, i.e., approximately the time interval corresponding to S1 and S2 labels. The magnetic trace for SPT and SUA, located in the dusk sector, shows an increase in *H*-component ranging from 50 nT to almost 100 nT at SC onset. However, larger enhancements of ~150 nT appear at IRT, which is located in the midnight sector. FRN and BSL, located in the pre-noon sector, show a negative change. Instead of a positive disturbance, a fast decrease takes place in both observatories leading to a negative depression similar to the Carrington event. However, a difference appears in this event: the magnetic trace in FRN shows a two-pulse structure with two negative *H*-spikes comparable in intensity.

At low latitudes, the SC appears at the four magnetic observatories involved in the computation of the *Dst* index. For these four observatories, as in the event on 29 October 2003, the disturbance of the two first hours following the SC depends strongly on the local-time sector. The horizontal component shows a negative depression at SJG reaching a minimum value of -52 nT, while the disturbance goes to positive values at KAK (in the post-midnight sector) reaching a value of 138 nT at the time of the maximum disturbance. HON and HER show also positive disturbance but reaching only less than 70 and 50 nT, respectively. The result of an average of all these measurements in the computation of the *Dst* index is the complete disappearance of the negative depression in SJG during the first hour after the SC. Indeed *Dst* index passed from -22 nT at 17 UT to 23 nT at 18 UT, remaining close to 20 nT for three hours. After 20 UT, *Dst* decreases steeply to less than -80 nT as a result of a decreasing horizontal component in all magnetic observatories involved in the index. The higher temporal resolution of *SYM-H* index does not prevent from missing the negative disturbance of the first hour after the SC, as it is the result of the average.

*Kozyra et al.* [2014] reviewed the unusual geospace consequences of the event on 21 January 2005. Although their study is mainly focused on the interval after the solar filament material arrival (~18:45 UT), it also includes interesting information regarding the first hours after the SC. Specifically, regarding the exceptionally intense auroral activity during this event, which is typical of superstorms, *Zhang et al.* [2008] reported dayside, duskside and nightside detached auroras on 21 January 2005 just after the fast shock at 17:12 UT with a life time of only 18 minutes. The duskside detached aurora, whose separation to the oval was not visible at the SI-12 camera, on board the IMAGE satellite [*Mende et al.*, 2003], started to expand towards nightside at 17:17 UT. The nightside detached aurora was observed around 55º MLat, well below the nightside edge of the auroral oval around 63º MLat.

The estimated hemispheric power index from POES in the Northern hemisphere was enhanced from 17:13 UT on 21 January until the end of the day, reaching up to 427.8 GW at 18:54 UT; these values are well related to the intense auroral activity described above. Aurora sightings between 19:00 and 19:30 UT were reported from The Netherlands (see for example http://shopplaza.nl/astro/pictures/NLIGHTS.HTM and http://www.ngc7000.org/photo/aurora20050121.html ), located at ~50º MLat.

## 4. Solar and interplanetary observations leading to 'Carrington-like' disturbances.

### 4.1. The triggers of the C03 event

The solar activity of late October and early November 2003 and its interplanetary consequences are one of the most analysed space weather events presented in a considerable number of research papers [see for example *Gopalswamy et al.*, 2005 and *Cid et al*., 2014 and references therein].

When comparing C03 and C59, solar features show some resemblance between both events. First of all, the main visual descriptive feature is a white-light flare in a very complex active region. *Cliver and Dietrich* [2013] and references therein classified recently the Carrington flare as an X45 (±5) flare, but on 28 October 2003 at 11:00 UT an X17.2 flare took place. The solar active region hosting the 28 October 2003 flare was also very complex, resembling the sunspot region reported by Carrington, with heliographic coordinates 8ºE, 16ºS, not far from those of the great flare in 1859. Similarly to the Carrington event, a magnetic crochet coincided with the solar flare on 28 October 2003 [*Villante and Regi*, 2008]. The size of the magnetic crochet recorded on the Greenwich and Kew geomagnetograms was used to estimate the flare soft X-ray peak intensity ranging from no less than X10 and more than X48 [*Cliver and Svalgaard*, 2004].

The extreme scenario for the instruments measuring interplanetary activity during these days can be compared to the extreme scenario during the C59 event for terrestrial magnetic observatories. In particular, the Solar Wind Electron Proton Alpha Monitor (SWEPAM) [*McComas et al*., 1998] on board ACE spacecraft was affected because of two different issues: (1) penetrating radiation from solar energetic particles led to high instrument background levels, causing the solar wind tracking algorithm to fail at times, and (2) for several of the highest-speed points on 29-30 October 2003, the high energy part of the solar wind exceeded the search mode energy range [*Skoug et al*., 2004]. As a result, solar wind data provided through the ACE Science Center (http://www.srl.caltech.edu/ACE/ASC/) for 29-30 October have a data gap that includes the interval of interest in this paper. However, magnetic field data from the Magnetic Fields Experiment (MAG) [*Smith et al*., 1998] are fully available for the event.

Figure 7 shows the interplanetary magnetic field (in the Geocentric Solar Magnetospheric, GSM coordinate system) on 29 October 2003, and the subsequent magnetospheric activity as seen by *SYM-H* index. The horizontal magnetic field component measured at THY has also been plotted in this panel although it will not be discussed in this section (we will come back to this figure in Section 5). Vertical solid line indicates the shock (S) identified by *Skoug et al.* [2004] at 05:58 UT that triggered the SC observed in the *SYM-H* index at 06:12 UT, indicating a delay of 14 minutes between the interplanetary trigger at ACE (S) and the response on the ground (SC), as measured by the *SYM-H* index. Then, to better compare the ACE data and the response on the ground, the plots showing ACE data are shifted 14 minutes with respect to the bottommost plot. According to *Skoug et al.* [2004], downstream of the shock the solar wind speed exceeded 1500 kms$^{-1}$ for a 6-hour period, with a peak value of 2240 kms$^{-1}$, and the proton temperature exceeded $10^7$ K. Proton density values were uncertain for this time.

The interplanetary magnetic field was highly enhanced (up to 63 nT) for about one hour following the shock. The $B_y$ component was largely negative (~-35 nT) until 06:33 UT when it turned suddenly to positive. $B_z$ was negative downstream of the shock with brief incursions to positive and at 06:27 UT decreased sharply to large negative values that quickly reached -53 nT. At 06:39 UT $B_z$ turned sharply northward for about 20 minutes reaching ~45 nT. After 06:56 UT $B_z$ was close to zero or even positive (~5 nT) until 07:19 UT, although this interval corresponds to a decrease of *SYM-H* index of about 120 nT. The two large southward turnings took place at the interplanetary shock at 05:58 UT (~30 nT in 3 min) and at 06:27 UT (~57 nT in 2 min). The indicated times correspond to the mid-point of the time interval of the sharp turning (vertical lines in Figure 7, labeled as S and T).

Regarding the intense long-lasting southward IMF $B_z$ which might be expected to trigger so large an *H*-spike due to an enhanced ring current, it is observed on 29 October from 17:30 UT until the end of the day, i.e. approximately half a day after the *H*-spike, and no cause can happen after its effect! However, this long-lasting $B_z$ can explain the main phase as seen by the *Dst* index (assuming Burton equation, *Burton et al.* [1975]), which peaks at -353 nT at midnight of 29 October.

### 4.2. The triggers of the C05 event

A GOES X7.1-class flare at 06:40 UT on 20 January 2005 and a halo CME observed on just one uncontaminated LASCO image at 06:54 UT are the observations of the main solar activity related to the C05 event. In this case, the intensity of the X-flare was not enough to produce a magnetic crochet as in the C03 and C59 events and the active region (NOAA 10720) was not a very complex active region. Nevertheless, AR10720 presented large flaring activity and produced several fast halo CMEs in the previous days. The fastest CME of cycle 23, which may have reached a speed of ~3000 km s$^{-1}$ close to the Sun [*Pohjolainen et al.*, 2007], originated in this active region.

At the time of C05 event, the *Dst* index was already disturbed after a moderate geomagnetic storm that happened two days before. Interaction between two CMEs and a high speed stream associated with a coronal hole characterized the interplanetary event. *Foullon et al.* [2007], *Rodriguez et al.* [2008] and *Kozyra et al.* [2013] provided a detailed study of the interplanetary signatures and the related solar activity for the event on 21-22 January 2005.

Figure 8 shows solar wind plasma and interplanetary magnetic field data for the C05 event. On 21 January 2005 at 16:46 UT, the ACE spacecraft detected the passage of an interplanetary forward shock (solid vertical line S1) that triggered the SC observed in the *SYM-H* index at 17:11 UT, indicating a delay of 25 minutes between the interplanetary trigger at ACE and the response on the ground, as measured by *SYM-H* index. As in Figure 7, to better compare the ACE data and the response on the ground, the horizontal axis of the plots showing ACE data in Figure 8 are shifted 25 minutes with respect to the bottommost plot.

The arrival of the S1 shock is followed by a secondary front discontinuity (solid vertical line S2) at 18:21 UT. While the interval after S1 shows very high temperature (~7×10$^5$ K) and densities of about 15 cm$^{-3}$, after S2 the plasma is cooler and the density is highly enhanced, peaking at ~60 cm$^{-3}$. After S1, the $B_z$ interplanetary magnetic field component is fluctuating, first pointing southward (16:54-17:22 UT) and then northward (17:22-17:51 UT). In the end it turns southward and smoother reaching ~-28 nT until S2. After this discontinuity, the magnetic field is northern almost all the time, although this interval corresponds to the main phase of the storm, as indicated by the decrease of *SYM-H* index. The horizontal magnetic field component measured at BSL has also been plotted in this panel although it will not be discussed in this section (we will come back to this figure in the next Section).

Our analysis of the two first hours after S1 reveals two sharp southward turnings of $B_z$ (shown as vertical dashed lines in Figure 8, labeled as T1 and T2) changing from northward to southward of about 30 nT in 2 minutes at 16:54 UT and at 17:52 UT (these times correspond to the mid-point of the time interval time of the sharp turning).

## 5. Discussion

When performing a systematic search for 'Carrington-like' profiles in the horizontal component recorded at mid- and low-latitude observatories spread in longitude, only two events arise whose temporal profiles during two days are extraordinarily similar to the one recorded at Colaba in 1859 (C59): the event on 29 October 2003 (C03) and the event on 21 January 2005 (C05). The similarities between the C03 event and the C59 event were first discovered by *Cid et al.* [2015]. Their results showed that although recorded magnetic field variations at one of the observatories (THY) closely resembled the profile recorded at Colaba during the Carrington event, a large asymmetry appeared in the disturbance when moving in longitude for C03. As a result, the disturbance of C03 in *SYM-H* and *Dst* indices was missed when averaging local disturbances.

As in C03, the magnetic disturbance of C05 is also missed by the *SYM-H* and *Dst* indices. When analysing at mid-latitude local magnetic records during the C05 event, a large asymmetry appears also in the disturbance when moving in longitude. As in the C03 event, in C05 the disturbance depends strongly on the magnetic local time. Even more, in both events similar profiles correspond to similar MLTs, being negative in the dayside and positive in the nightside: peaking at dawn-noon sector and almost unnoticed in the noon-dusk sector. In both events less-than-one-hour local depressions (*H*-spikes) happened simultaneously with auroral effects, with auroral displays observed at locations up to 10º away in latitude from the stations that recorded the maximum magnetic disturbance. This scenario, where a clear day-night asymmetry appears when moving in longitude, is indicative of a disturbance caused by R1 FAC [*Yu et al.*, 2010]. Indeed, according to the Biot-Savart's law, the horizontal geomagnetic field ground-disturbance ($\Delta H$) at mid latitude due to R1 FACs is negative in the dayside and positive in the nightside, in line with what is observed in both C03 and C05 events. Moreover, the only available record at low latitude for C59, which fits to C03 and C05 patterns at the same MLT, supports that FACs are the main magnetospheric current system involved in Carrington-like disturbances, as previously suggested by *Cid et al.* [2015]. On the other hand, when applying the Biot-Savart's law to an enhanced ring current, a depression in $\Delta H$ is expected at any longitude, and as a result at any MLT, making low-latitude global indices such as *Dst* or *SYM-H* appropriate to study magnetic disturbances related to a symmetric ring current enhancement, but making them unsuitable when large asymmetries (positive and negative disturbances) appear in longitude. Indeed, the *ASY-H* index was introduced to describe the longitudinally asymmetric geomagnetic disturbance field at mid latitude due to a partial ring current enhancement, which results in a dawn-dusk asymmetry. In the case of R1 FACs, due to the day-night asymmetry at mid and low latitude, those indices are unsuitable to quantify the disturbance.

To reinforce the hypothesis of FACs as the main magnetospheric current system involved in Carrington-like disturbances, a deeper analysis has been carried out with the subsets of events from Figure 5 with membership to group C (C-score > 0.85 and B-score < 0.85) and with those assigned to group B (C-score < 0.85 and B-score > 0.85). Figure 9 shows a scatter plot of score versus MLT for each subset (group-C events appear as red crosses and group-B ones in green) for high-, mid- and low- latitude ground geomagnetic stations.

Looking at the bottom panels (a) and (b) in Figure 9, the distribution in longitude of group B events at low and mid latitude shows a clear dawn-dusk asymmetry. Indeed, $\Delta H < 0$ appears almost at any MLT (indicative of a ring current enhancement), with the largest scores peaking at dusk sector as indicative of the major role of the partial ring current in the initial phase of typical geomagnetic storms.

On the other hand, in panels of Figure 9c, a day-night asymmetry in the distribution of events (with most of the events at night sector) appears at high latitude for both group B and group C, as indicative of the significant occurrence of substorms of the tail current system at latitudes close to the auroral electrojet. It should be also noticed the highest score values correspond to group C at high latitude.

A day-night asymmetry appears again at mid latitude for group C, but in this case the *H*-spike of Carrington-like events concentrate in day-time, departing from the substorms affinity with night-time. Despite the small number of events in the plot, the large difference to the equivalent figure at mid latitude for group B is enough to suggest another current as cause responsible for the magnetic disturbance.

The MLT distribution of the highest C-score events at mid latitude points out to R1 FACs as the most probable cause of C-events. Nevertheless, the analysis of the evolution of the magnetic disturbance in longitude for every event is essential to conclude the direct effect through the Biot-Savart's law of R1 FACs as the cause of the *H*-spike disturbances for the Carrington-like events.

Searching for the interplanetary trigger of C03 and C05, we find out that in both events the drop of the *H*-spike (S and T in Figure 7, T1 and T2 in Figure 8) took place when southward incursions of the IMF larger than 10 nT/min under high dynamic pressure (note that the horizontal axes in bottom panel in Figures 7 and 8 are shifted 14 min for C03 and 25 min for C05 in order to make coincident the interplanetary shock and the SC). Indeed southward abrupt reversals have been identified as the cause of prompt electric field penetration [*Othani et al.*, 2013], which is favoured by high solar wind density/pressure [*Lopez et al*., 2004; *Fiori et al*., 2014]. The end of each drop in *H* (the peak of the spike) is well correlated to a northward IMF turning as well. These results suggest that the disturbances recorded during C03 and C05 were controlled by the magnetic reconnection process. Similar results were obtained by *Wei et al.* [2008] when analysing a long-lasting penetration event observed during 11-16 November 2003. Southward turnings instead of long-lasting and intense southward IMF $B_z$ were considered as better solar-wind precursors of large variations of *Dst* [*Saiz et al.*, 2008]. *Skoug et al.* [2004] also pointed out that large alternating northward and southward IMF contributed to extreme geomagnetic storms as 4 August 1972, 15 July 2000, and 31 March 2001.

*Tsurutani et al.* [2003] proposed that the most likely driver for the *H*-spike in C59 (what they called the main phase of the Carrington storm) would be a magnetic cloud due to its long duration and intense southern $B_z$, but no interplanetary data were available at that time to check this hypothesis. Following this hypothesis, the Carrington event was associated with very extreme solar wind and IMF conditions in order to reproduce a *Dst* value of -1600 nT [*Li et al.*, 2006] or a ground-magnetic disturbance of the same value at one location, i.e. Colaba [*Ngwira et al.*, 2014]. Their models yielded an extremely large IMF $B_z$ component (> -200 nT for 2.5 hours) and an equally large density (400-800 cm$^{-3}$) and velocity (~ 2000 km s$^{-1}$). Although coronagraph images show CMEs travelling at speeds like that one assumed by *Ngwira et al.* [2014] however the magnetic field and the solar wind density proposed are unprecedented: the strongest absolute field strengths reported at 1AU are 60-80 nT (less than one third) and the largest observed density is below 200 cm$^{-3}$.

In the cases of C03 and C05, the driver was the sheath downstream of the interplanetary shock, where the abrupt reversals of IMF $B_z$ are combined with a high dynamic pressure. Solar wind pressure has been pointed out above as a factor which favours prompt electric field penetration, as it produces a larger compression of the magnetosphere, which results in larger FACs and as a consequence, in a larger peak at the spike in the pre-noon sector.

In spite of the lack of solar wind data in some of the events studied in this paper, the time of transit of all of them is well established: ~17.5 h for C59 [*Tsurutani,* 2014], ~19 h for C03 [*Skoug et al.,* 2004] and ~34 h for C05 [*Pohjolainen et al.*, 2007]. A smaller transit time is equivalent to a larger solar wind velocity and therefore larger dynamical pressure will be expected. The observed peak values at the *H*-spike in the three Carrington-like events (C59, C03 and C05) agree with these expectations.

Even more, assuming by extrapolation that the trigger of C59 was also an abrupt reversal of IMF $B_z$ combined with a high dynamic pressure, a linear extrapolation can be performed to estimate solar wind conditions preceding the Carrington event. As the intensity of the *H*-spike in C59 is 2.3 times that of C03, this factor let us estimate a solar wind density close to 230 cm$^{-3}$, an abrupt southward reversal of IMF $B_z$ of about 135 nT/min and later on a southward turning with a long-lasting $B_z$ value of about -50 nT for about 7 hours. Similar values are obtained by extrapolation of solar wind data conditions during C05. According to historical solar wind data records, these values are much more reasonable than those previously considered in simulations. Nevertheless, the profile of the disturbance resulting at ground level is expected to be highly dependent on longitude (as in C03 and C05), contrary to the results shown in Figure 9 from *Ngwira et al.* [2014] where a similar pattern appears at low-latitude ground locations at different longitudes.

## 6. Conclusions

We have conducted a systematic search for 'Carrington-like' profiles in local magnetic records at high, mid and low latitude on the period from 1991 to 2009. At mid and low latitude the search results only in two events: the event on 29 October 2003 at Tihany magnetic observatory in Hungary (C03) and the event on 21 January 2005 at Stennis magnetic observatory in southern US (C05). The similarities between C03 and C05 and their extrapolation to C59, led us to the following results:

1) As previously suggested by *Cid et al.* [2015] for C03 and C59, our analyses based on ground-magnetometers records of C05 at low- and mid-latitude observatories, and on the similarity of the temporal profiles observed for these events at different MLTs, support the suggestion of *Cid et al.* [2015] that FACs play a major role in the large *H*-spike of C05 and for all Carrington-like events observed at mid latitude, discarding the role of the ring current as the major current involved in this kind of disturbance.

2) As observed in C03, also in C05 the *H*-spike is missed in indices such as *Dst* or *SYM-H*. This fact arises due to the large differences appearing among the local magnetic disturbances from observatories spread in longitude (positive in some and negative in others) and averaged out. Besides the large differences in local magnetic records at the same magnetic latitude in every event, similar profiles are recorded for C03 and C05 at similar MLTs. A similar pattern appears also in C59 at the only available MLT at low latitude (the one recorded at Colaba). Thereby, as a common result for all Carrington-like events, it reinforces the awareness about the possibility of missing hazardous space weather events such as the large *H*-spike recorded at Colaba in September 1859 by using global geomagnetic indices.

3) The non-occurrence of any Carrington-like event at low latitude in the period analysed reinforces the uniqueness of the Carrington event recorded at Colaba. But at the same time, our results evidence a greater probability to observe *H*-spikes at mid latitudes than expected according to previous analyses based on global indices like *Dst*.

Furthermore, the occurrence or non-occurrence of an *H*-spike event at mid-low latitude seems to be no dependent on the intensity of the storm as seen by *Dst* or *SYM-H* index.

4) Large southward abrupt reversals of IMF and high solar wind pressure appear as the interplanetary trigger of the *H*-spike of events C03 and C05. Both signatures, which appear in the sheath, are associated with prompt electric field penetration. Extrapolation of solar wind data observed conditions during these events provides driving conditions for C59 which are not unusually high attending to measurements available at 1 AU.

The findings described in this paper not only shed light on the intensity, triggers and currents involved during the Carrington storm, but also may discard some well-established theoretical framework beneath the common understanding of Sun-Earth interaction such as the local character of the ground disturbance, which diminishes the value of the *Dst* for the Carrington storm, the relevance of southward reversals of IMF as triggers of ground disturbances and the role played by some magnetospheric currents commonly forgotten in empirical approaches and simulations to analyse the still unique

Carrington storm. A systematic analysis of southward IMF reversals under large pressure and their geospace consequences is nowadays under development.

The solar source of abrupt southward IMF reversals should be also identified in order to forecast local large geomagnetic disturbances as *H*-spikes of Carrington-like events. An accurate forecasting of local magnetic field disturbances as *H*-spikes should be a major target in the risk scenario identified as a threat to vital infrastructure such as power grids.

**Acknowledgements**. Authors are grateful to the principal investigators and teams of the SWEPAM and MFI experiments on ACE, and NOAA/POES. Authors thank teams for data from GOES and also from LASCO catalogue. Geomagnetic field data have been obtained from INTERMAGNET magnetic observatories. The authors thank the Institutes that operate the observatories which provided data for this study; they also acknowledge the WDC for Geomagnetism for the magnetic field data and the *SYM-H* index.

This research was supported by the contract 'Estudio de la influencia de fenómenos relacionados con la Meteorología Espacial en las infraestructuras de REE' between Red Eléctrica de España and the Universidad de Alcalá and by grants PPII10-0183-7802 from the Junta de Comunidades de Castilla-La Mancha of Spain and AYA2013-47735-P from MINECO.

The editor thanks Mauro Regi and an anonymous referee for their assistance in evaluating this paper.

**Table 1.** Location of INTERMAGNET ground stations used in this work. Geomagnetic coordinates are given by using the IGRF geomagnetic field model year 2000

| Observatory | IAGA code | Geodetic latitude (deg) | Geodetic longitude (deg) | Magnetic latitude (deg) | Magnetic longitude (deg) |
|---|---|---|---|---|---|
| Narsarsuaq | NAQ | 61.16 | 314.56 | 69.90 | 38.41 |
| Fort Churchill | FCC | 58.76 | 265.91 | 67.93 | 327.91 |
| Abisko | ABK | 68.36 | 18.82 | 65.89 | 114.8 |
| Tihany | THY | 46.90 | 17.54 | 45.87 | 100.06 |
| Fresno | FRN | 37.09 | 240.28 | 43.42 | 304.93 |
| San Pablo | SPT | 39.55 | 355.65 | 42.69 | 75.88 |
| Toledo | SUA | 44.68 | 26.25 | 42.38 | 107.50 |
| Surlari | IRT | 52.27 | 104.45 | 41.68 | 176.8 |
| Irkutsk | BSL | 30.35 | 270.36 | 40.06 | 339.49 |
| Stennis S. Center | SJG | 18.11 | 293.85 | 28.2 | 6.02 |
| San Juan | KAK | 36.23 | 140.18 | 26.99 | 208.50 |
| Kakioka | HON | 21.32 | 202.0 | 21.5 | 269.48 |
| Honolulu | | | | | |
| Alibag | ABG | 18.62 | 72.87 | 9.91 | 145.96 |
| Hermanus | HER | -34.43 | 19.23 | -33.75 | 83.73 |

**Figure 1.** Frequency distributions of the Carrington-like scores (indicating similarity with Carrington event) of the events registered on three geomagnetic observatories at high (FCC), mid (THY) and low (ABG) latitude.

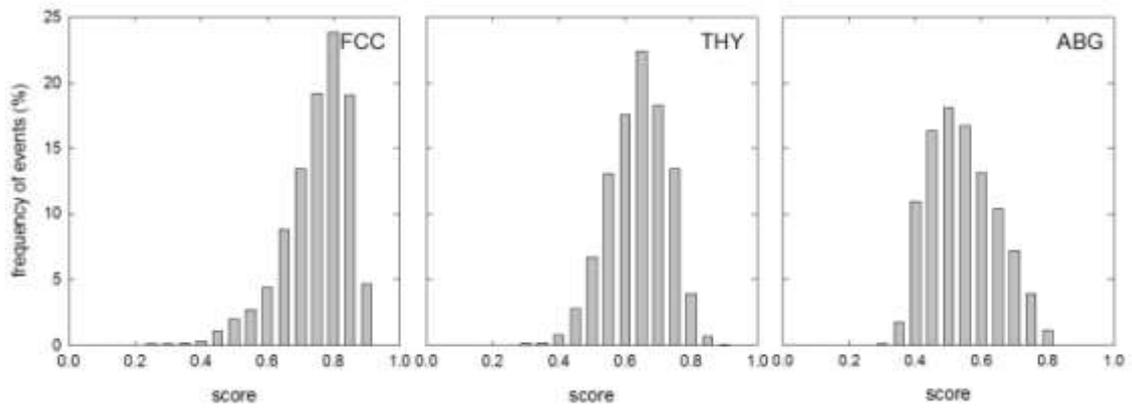

**Figure 2.** Highest score events at high- (panel a), low- (panel b) and mid- (panels c and d) latitude observatories, FCC, KAK, THY and BSL respectively. Normalised $H$ component of every observatory is shown in black and superposed is Model-C (red), which corresponds to normalised Carrington measurements.

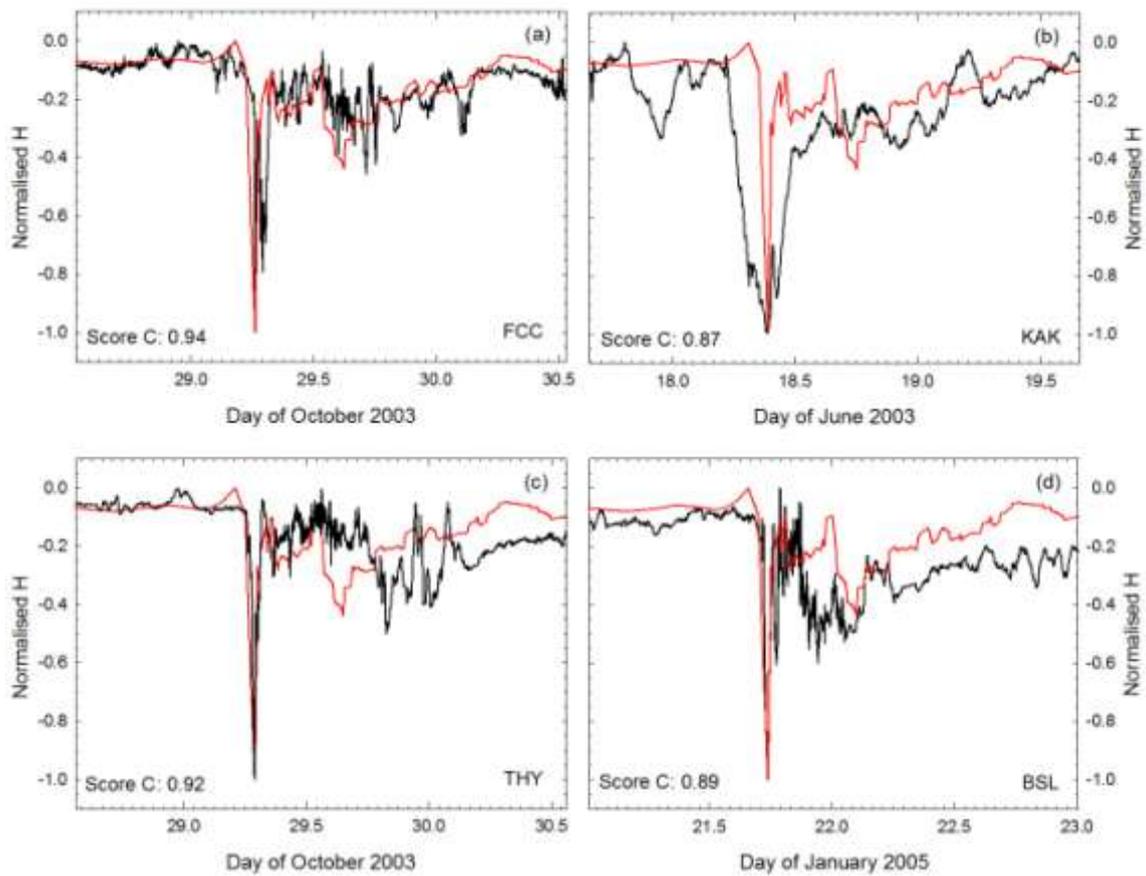

**Figure 3.** Flowchart of the method adopted for cluster analysis of Carrington-like and Basic-like (typical) geomagnetic events.

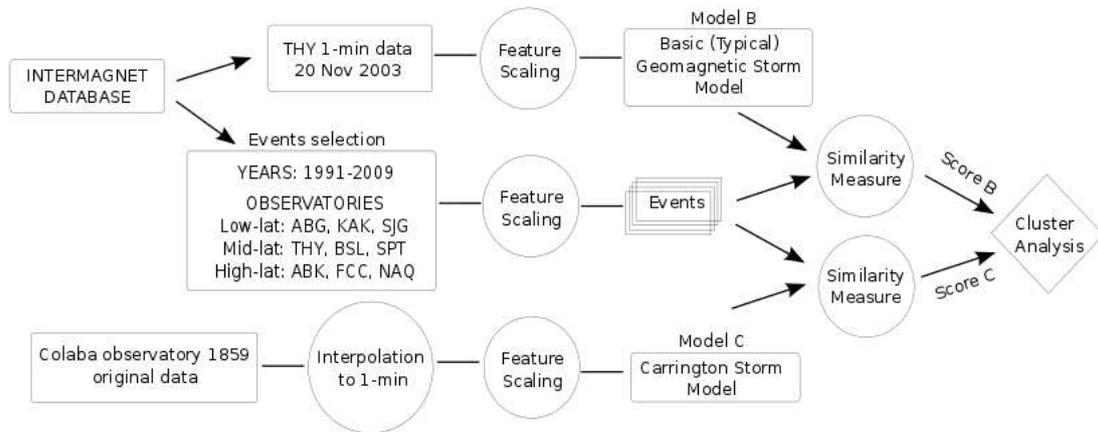

**Figure 4.** 48-h time window corresponding to the magnetic disturbance of B-Model (green) and C-Model (red) used by CLSE in order to obtain the event list.

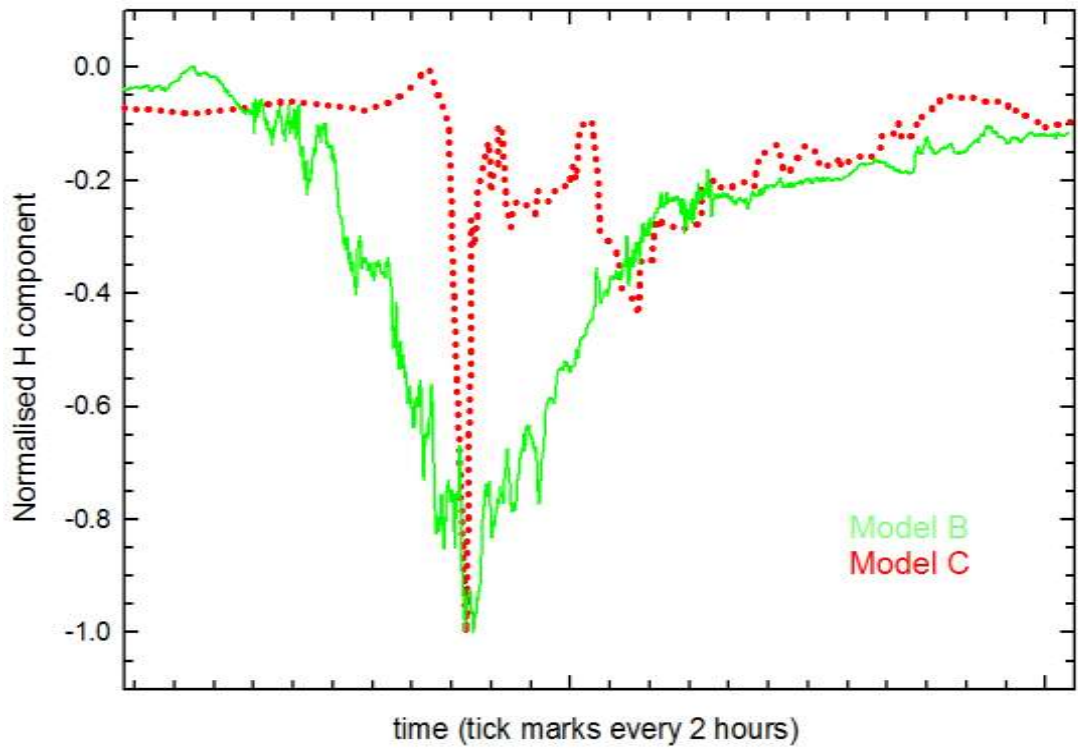

**Figure 5.** Representation used for cluster analysis of the scores (Score B and Score C) obtained from similarity measures with Model-B (Basic) and Model-C (Carrington-like), respectively. Panel (a) includes results from ABG, KAK, and SJG (low latitude), panel (b) THY, BSL, and SPT (mid latitude) and panel (c) ABK, FCC, and NAQ (high latitude). C03 and C05, selected at panel (b) for their location in group C (see Section 2.2), correspond to 29 October 2003 and 21 January 2005 events.

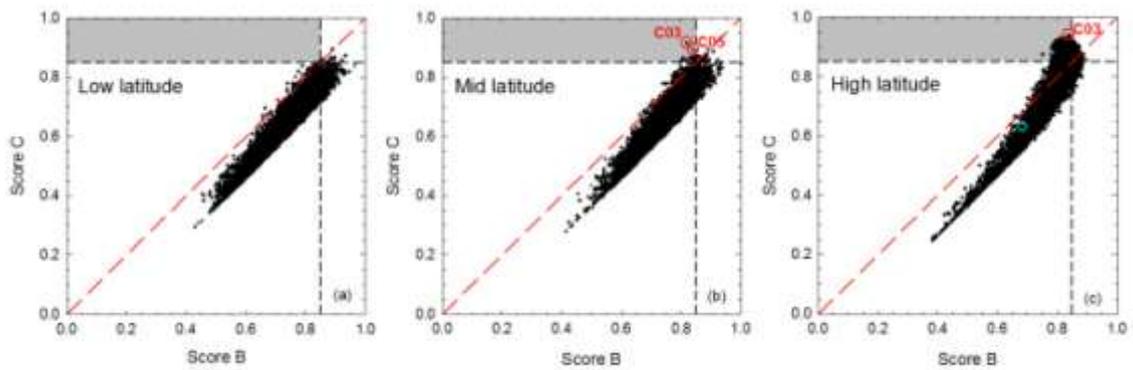

**Figure 6.** Geomagnetic disturbances of the C05 event (21 January 2005). Panel (a) shows measurements of the *H* component recorded at mid-latitude observatories. From top to bottom: SUA, IRT, FRN, BSL, and SPT (increasing in MLong). Panel (b), from top to bottom, displays *SYM-H* and *Dst* indices (top panel), and the *H* component recorded at the four low-latitude observatories involved in the computation of the *Dst* index: HER, KAK, HON, and SJG. In every upper-right corner, the subtracted magnetic field value and the difference between the maximum and the minimum value in the last 24 hours preceding the disturbance ($\Delta H_{\text{max 24h}}$) have been indicated. Vertical solid lines, S1 and S2, mark shock discontinuities and vertical dashed lines T1 and T2 mark southward turnings of $B_z$. These marks are also shown in Figure 8 to facilitate further analysis.

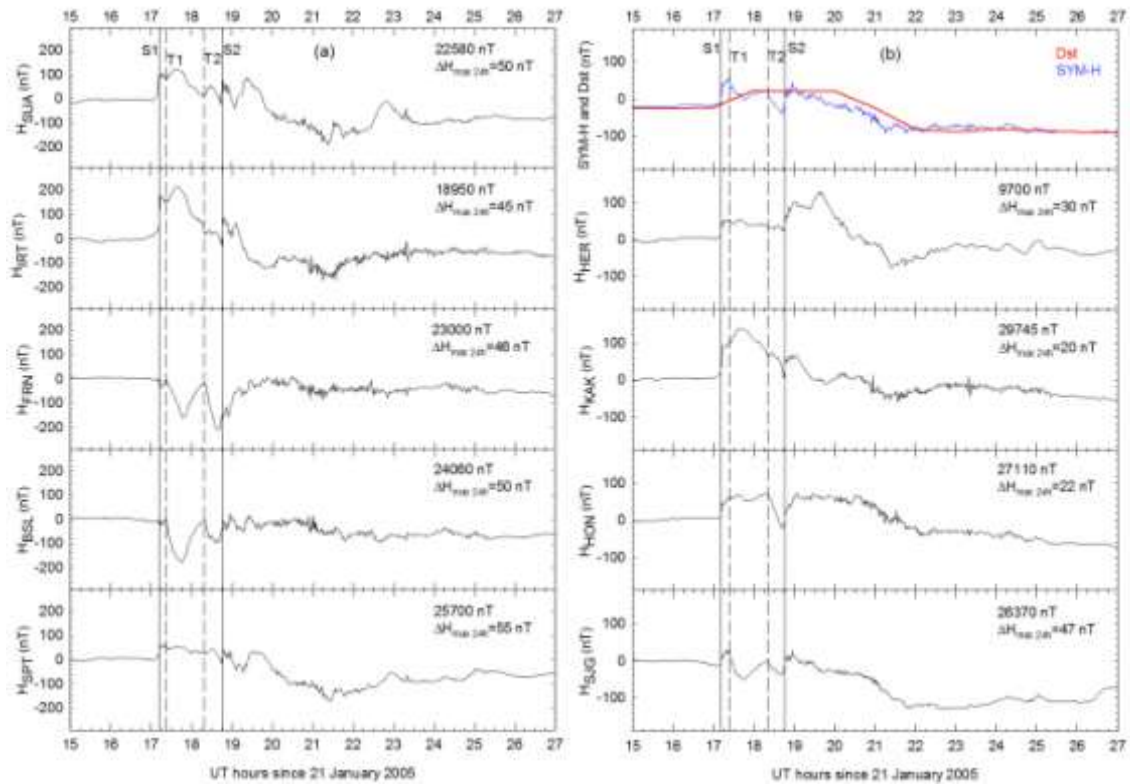

**Figure 7.** Interplanetary and geomagnetic data for C03 event (29 October 2003). From top to bottom: interplanetary magnetic field measured by ACE spacecraft at L1 point (4 top panels, magnitude and XYZ GSM components). Bottom panel shows the magnetic disturbance at the Earth's surface (*SYM-H* index and THY observatory). Two different time-lines in horizontal axis have been set: top labels correspond to ACE data (four top panels) and bottom labels correspond to ground data (bottom panel). Interplanetary plots are shifted 14 minutes (see text for details) relative to ground plots. Vertical solid line S marks shock discontinuity. Vertical solid line S and dashed line T mark southward turnings of $B_z$.

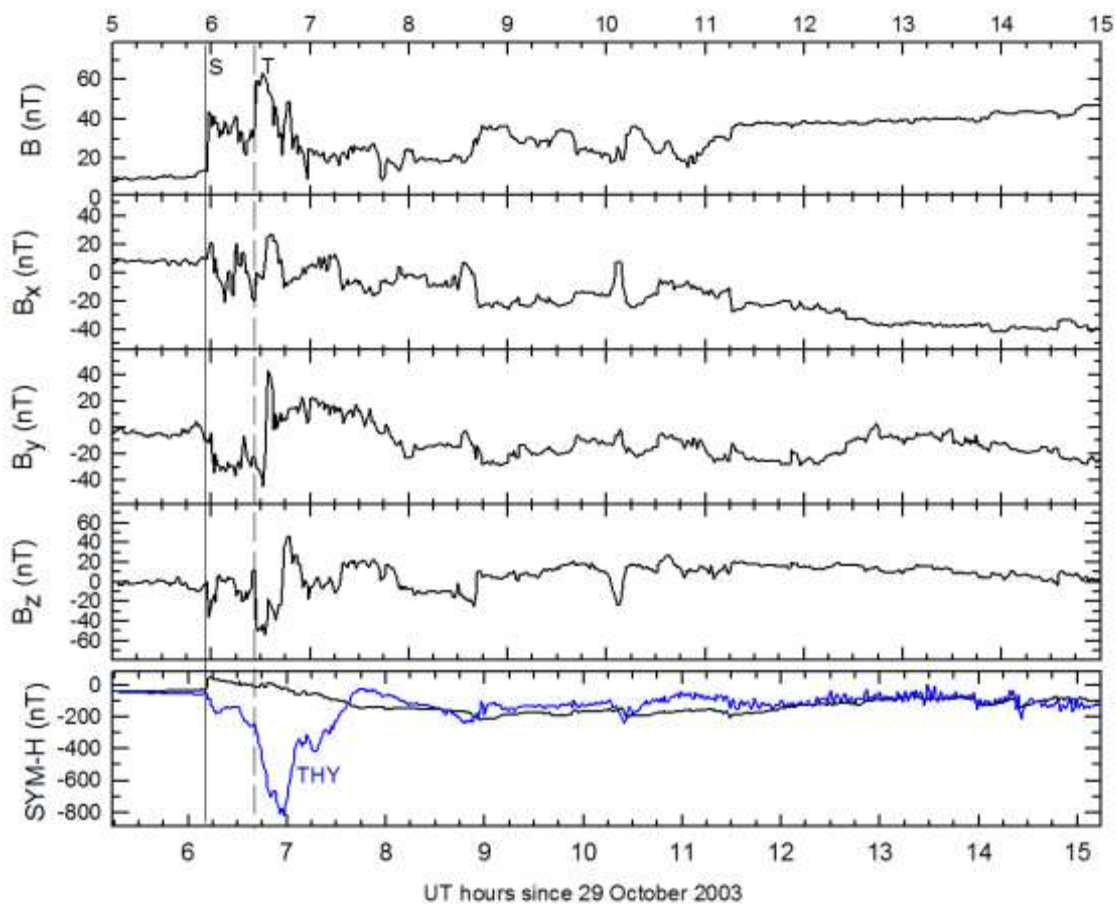

**Figure 8.** Interplanetary and geomagnetic data for C05 event (21 January 2005). Interplanetary magnetic field (same panel distribution as 4 top panels of Figure 5) and solar wind parameters, *Vsw*, *N*, *T,* in the next 3 panels, measured at ACE location. Eighth panel shows the dynamic pressure $P_{dyn}$. Geomagnetic data (*SYM-H* index and BSL observatory) are shown in bottom panel. Two different time-lines in horizontal axis have been set: top labels correspond to ACE data (eight top panels) and bottom labels correspond to ground data (bottom panel). Interplanetary plots are shifted 25 minutes (see text for details) relative to ground plots. Vertical solid lines, S1 and S2, mark shock discontinuities and vertical dashed lines T1 and T2 mark southward turnings of $B_z$.

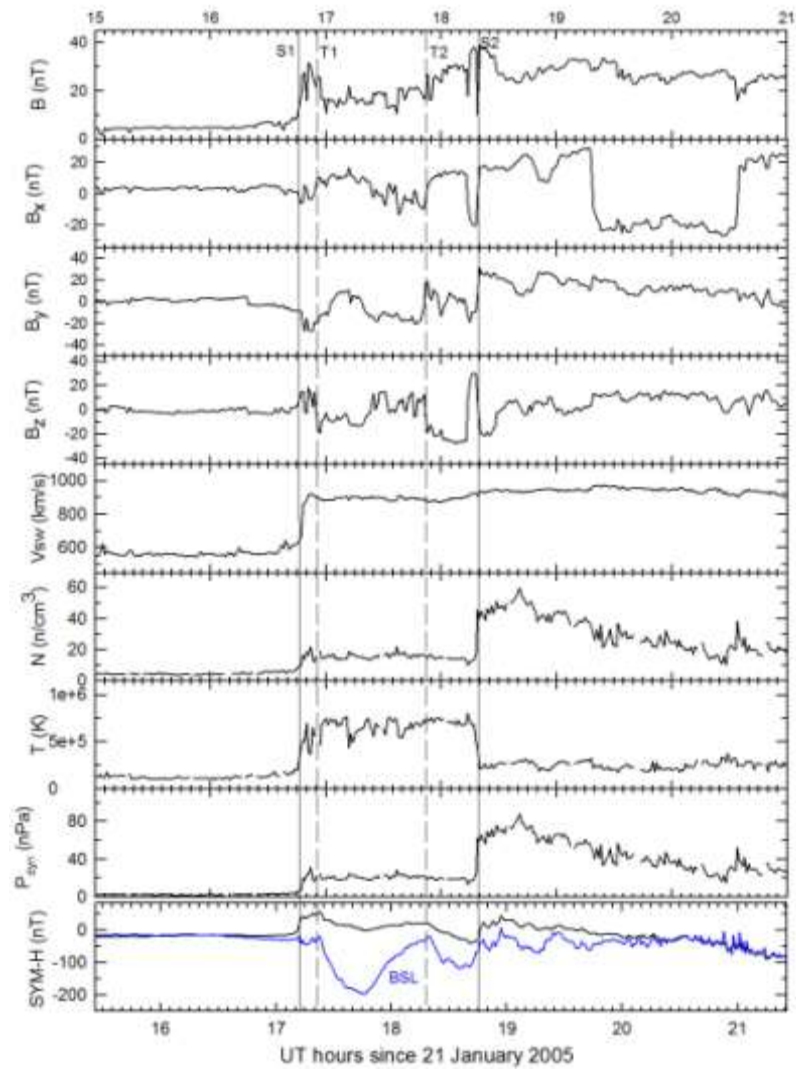

**Figure 9**. Event score distribution with respect to the MLT for high- (a), mid- (b) and low- (c) latitude geomagnetic stations. Red plus signs in upper panels correspond to Carrington-like events [group C], and green in bottom panels correspond to group-B events. In upper panel (b) two red circles denote location of the C03 (29 October 2003) and C05 (21 January 2005) events.

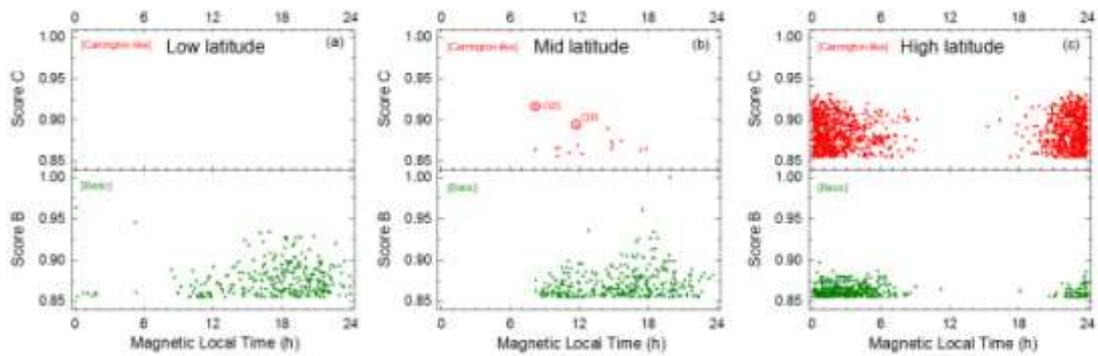